\documentclass{article}
\usepackage{graphicx}
\usepackage{amsmath}

\begin{document}
\title{Finite Temperature Tunneling and Phase Transitions in
$SU(2)$-Gauge Theory}

\author{A. V. Shurgaia\thanks{On leave from A. Razmadze Mathematical
Institute of the Georgian Academy of Sciences, 0193 Tbilisi, Georgia.
Email: avsh@rmi.acnet.ge },  H. J. W. M\"uller-Kirsten\thanks {Email:
mueller1@physik.uni-kl.de},  J.-Q. Liang\footnote{Department of Physics
and Institute of Theoretical Physics, Shanxi
 University  Taiyuan, Shanxi 030006, China. Email:
 jqliang@mail.sxu.edu.cn} \\ and D. K. Park\thanks
 {Department of Physics, Kyungnam University, Masan, 631-701,
Korea. Email: dkpark@genphys.kyungnam.ac.kr}} \maketitle
 \begin{center}
Department of Physics, University of Kaiserslautern,\\
D67653 Kaiserslautern, Germany
\end{center}

\begin{abstract}

A pure Yang-Mills  theory extended by addition of a quartic term is
considered in order to study the transition from the quantum tunneling
regime  to that of classical, i.e. thermal, behaviour. The periodic field
configurations are found, which interpolate between the vacuum and
sphaleron field configurations. It is shown by explicit calculation that
only smooth second order transitions occur for all permissible values of
the parameter $\Lambda$ introduced with the quartic term.
 The theory is one of the rare cases which can be handled analytically. \\

\end{abstract}

\section {Introduction}

One of the amazing phenomena of  quantum physics is the barrier
penetration due to tunneling processes. The occurrence of such processes
in different areas of physics (solid state physics, high energy
multiparticle scattering with baryon number violation, low-temperature
physics, nuclear reactions, the formation of the Universe
 etc.) does not leave any doubt that they do actually take place. The theory of
 tunneling has been studied in many ways. It has become evident that such processes
 are due to classical configurations, which are solutions of the classical equation
 of motion with Euclidean time, namely stable vacuum configurations, now called  instantons,
 which are responsible for transitions between topologically distinct vacua (relevant in
 explaining the high energy multiparticle collisions accompanied by baryon number violation)
 \cite{G.'t,Bel,Mat,Tin} and  unstable periodic configurations called periodic instantons
and periodic bounces, which determine the decay of  metastable physical
systems\cite{{Col1,Col2, Lan}}. On the basis of the latter a theory of
barrier penetration at finite energies has been
developed\cite{Aff,Lar,Gra,Zwa,Rei}.
  In particular it was shown, that the transition from temperature assisted
  tunneling to thermal activation can be considered as a phase transition which
  takes place as the temperature (or energy) of the system increases.  Dissipative
  forces do not affect the general features of  transitions. At zero temperature
  the barrier penetration is determined by tunneling with a rate  controlled by
  vacuum instantons and is proportional to $\exp (-S)$ where $S$ is the action.
  As the temperature increases the tunneling process   (temperature assisted
  tunneling) begins  to be suppressed and at sufficiently high energies (comparable
  with the height of the potential barrier) the system overrides the barrier (by
  thermal activation) and the penetration is governed by the Boltzman factor
  $\exp(-E_0/{kT})$ where $E_0$ is the energy of the system, corresponding to a
  particle sitting at the top of the barrier (the sphaleron). The configurations,
  which interpolate between these two processes are the periodic instantons
  \cite{Har,Kle,Lia1}. It was recently discovered, that depending on the shape of
  the potential barrier a transition of the first order  (i.e. a sharp transition)
  is also possible\cite{Chu1}. In the context of the Higgs model with some effective
  potential this type of transition was known before\cite{Lin}. In refs.
  \cite{Gor,Mue}  criteria for the occurrence of the transition of the first order
  have been derived  and examined for various quantum mechanical models.
  Furthermore it has been shown\cite{Kuz,Hyun}, that the periodic instantons may
  have
bifurcations, which qualitatively change the behaviour of the phase
transitions at finite temperature.

Recently large-spin systems turned out to be of increased interest, as
these exhibit the first order  phase transitions\cite{Chu2,Lia2}. The
specific feature of these systems is a nonlinearity of the kinetic term.
This could hint  at the existence of sharp first  order transitions  in
$\sigma$-models. One should keep in mind, that although the general theory
of tunneling is well understood , only few examples are known, which make
it possible to analyze the problem of phase transitions by explicit
analytical calculations. In field theory the problem is even more
complicated. In spite of these drawbacks  one has succeeded to investigate
the problem  in some field-theoretical models numerically or by reducing
the problem to a quantum mechanical
one~\cite{Hab,Par1,Rub,Par2,Lia3,Lia4,Gar,Fer,Hun,Par3,Par4}. Recent
investigations in scalar theories in $D+1$- dimensions with
$D=1,2,3$\cite{Gar,Fer,Hun,Par3} have shown the existence of both types of
phase transitions depending on the value of the parameter, which violates
a certain symmetry of the theory. From the point of view of the
electroweak processes the $\sigma$-model and  Abelian-Higgs model in two
dimensions are of more interest, since as toy models they exhibit some
features of the electroweak theory. It has been
shown\cite{Hab,Par1,Par2,Par3,Par4} that whereas the $\sigma$-model with
the symmetry-breaking term allows the first-order phase transition (adding
the Skyrme-term allows the second-order transition too\cite{Par1}), the
Abelian-Higgs model admits only second-order phase transitions. It is
interesting, that in the model the sharp first-order transition takes
place, if the spatial coordinate is compactified\cite{Par4}. Recently
complex periodic instantons have been  investigated in
$SU(2)$-Yang-Mills-Higgs theory\cite{Bon}. New classes of bifurcating
periodic instantons of the classical equations of motion were found
numerically, depending on the self-coupling constant of the scalar field.

In the present work we investigate a theory with a vector field, namely
globally $SU(2)$-invariant theory. The study of vector fields is, in our
opinion,  of interest from the point of view of   such a complicated
theory as the electroweak theory. The theory we are interested in is  a
pure Yang-Mills theory extended by addition  of a quartic term. This term
violates the local gauge invariance of theory,
 but there is still a global invariance. Although adding this term will destroy
  renormalizability this is irrelevant to the problem we are going to
  investigate. Considering the model in the
  framework of the field theory constructed on the sphere $S^3$ one succeeds to
  reduce the problem to the quantum mechanical one with the effective potential
  which is an asymmetric double-well potential and depends on the extra parameter,
   introduced with the quartic term.  We show, that
for all values of this parameter  transitions are of second-order. The
model we consider here is one of very few in more than two dimensions
which can be treated analytically
 and therefore deserves particular attention even in spite of its idealization.

\section {The model}

The model we consider is described by the Euclidean action
\begin{equation}
S=\int d^4x\left\{\frac{1}{4}F_{\mu \nu a}F_{\mu \nu a}-\frac{\Lambda
g^2}{12}(A_{\mu a} A_{\mu a})^2\right\}
\end{equation}
in which as usual
\begin{equation}
F_{\mu \nu a}=\partial_\mu A_{\nu a} -\partial_\nu A_{\mu
a}+g\varepsilon_{abc}A_{\mu b} A_{\nu c}.
\end{equation}
Although the additional quartic term breaks the local gauge invariance the
model is still  globally $SU(2)$-invariant   and   is of interest from a
field-theoretical point of view. The model admits
pseudoparticle-antipseudoparticle classical field
configurations\cite{Dol}. We present below  periodic field configurations,
which are responsible for quantum-classical phase transitions. In what
follows we shall work in the framework of a field-theoretical approach
\cite{Fub},  which is constructed on the sphere $S^3$ embedded in a
4-dimensional Euclidean space. This approach is  especially convenient for
conformally invariant theories, as the system can be considered to evolve
along the radius of $S^3,$ in which case the operator of scale
transformations  becomes an evolution operator. Thus the radius $r$,
namely the parameter $\sigma =\ln r,$ is a proper time of the physical
 system and the operator of  scale transformations is considered as the ``scaled
 energy'' of the system.

We shall illustrate this by considering the simplest example of the scalar
field $\Phi(x)$. The general conformal group contains a dilatation
operator $D$  defined in terms of the field $\Phi(x)$ as
$$
D=\int d^3x x_\nu T_{4\nu}
$$
where $T_{\mu \nu}$ is the energy-momentum tensor. The scaling
transformation for the field is
$$
i[D,\Phi(x)]=x_\mu\partial_\mu\Phi(x)+\Phi(x).
$$
We  introduce spherical coordinates in 4-dimensional Euclidean space by
setting
\begin{equation}
x_\mu = rn_\mu
\end{equation}
with a unit vector
\begin{equation}
n_\mu  =(\sin{\psi}\sin{\theta}\cos{\phi},\,
\sin{\psi}¸\sin{\theta}\sin{\phi}, \,\sin{\psi}\cos{\theta},\,\cos{\psi}),
\end{equation}
and define the field $\chi(r,n_\mu)=r\Phi(x)$. In terms of the new
coordinates and the field $\chi(r,n_\mu)$ the transformation law reads:
$$
i[D,\chi(r,n_\mu)]=\frac{\partial \chi(r,n_\mu)}{\partial \ln r},
$$
which makes the role of $D$ as the evolution operator evident. One can
also find the following integral representation for D:
$$
D=\int d\Omega\left\{\frac{\partial\mathcal
L}{\partial\partial\chi/\partial\ln r}\frac{\partial\chi}{\partial\ln
r}-\mathcal L\right\},
$$
which is the Legendre-transform of the Lagrangean $\mathcal L$ integrated
over the angles. The action of the system is then
$$
S=\int d\ln r d\Omega \mathcal L.
$$
Thus we conclude, that in field theory constructed on $S^3$ the energy is
replaced by the eigenvalues of the operator of the scaling
transformations. The temperature is defined as the inverse of the period
of periodic field configurations expressed in terms of the proper time
$\ln r$. The period is determined as the derivative of the action with
respect to the ``scaled energy''.

We now return to our model. We look for periodic field configurations of
period $P({E})$, where $E$ is a ``scaled energy'' of the system, which
interpolate between the vacuum and the unstable field configuration named
sphaleron as the proper time $\tau$ varies from $-P({E})/2$ to
$P({E})/2.$ We  make  the Ansatz
\begin{equation}
A_{\mu a}=\frac{1}{g} \eta_{a\mu \sigma } n_{\sigma} \frac{\phi(r)}{r}
\end{equation}
with the spherically symmetric function $\phi(r)$ to be determined.
Performing the integration  frac12the angle variables we obtain (with
$\tau =\ln r$):
\begin{equation}
S=\frac{6\pi^2}{g^2}\int_{-P({E})/2}^{P({E})/2} d\tau
\left\{\frac{1}{2}\left({\frac{\partial\phi(\tau)}{\partial\tau}}\right)^2+
V(\phi(\tau))\right\},
\end{equation}
in which  $V(\phi(\tau))$ is an effective potential in terms of the
function $\phi(\tau)$:
\begin{equation}
V(\phi(\tau))=\frac{(2-\Lambda)}{4}\phi^4(\tau)-2\phi^3(\tau)+2\phi^2(\tau)
\end{equation}
The shape of the potential depends on the parameter $\Lambda$.
\begin{center}

  \includegraphics[width=8cm]{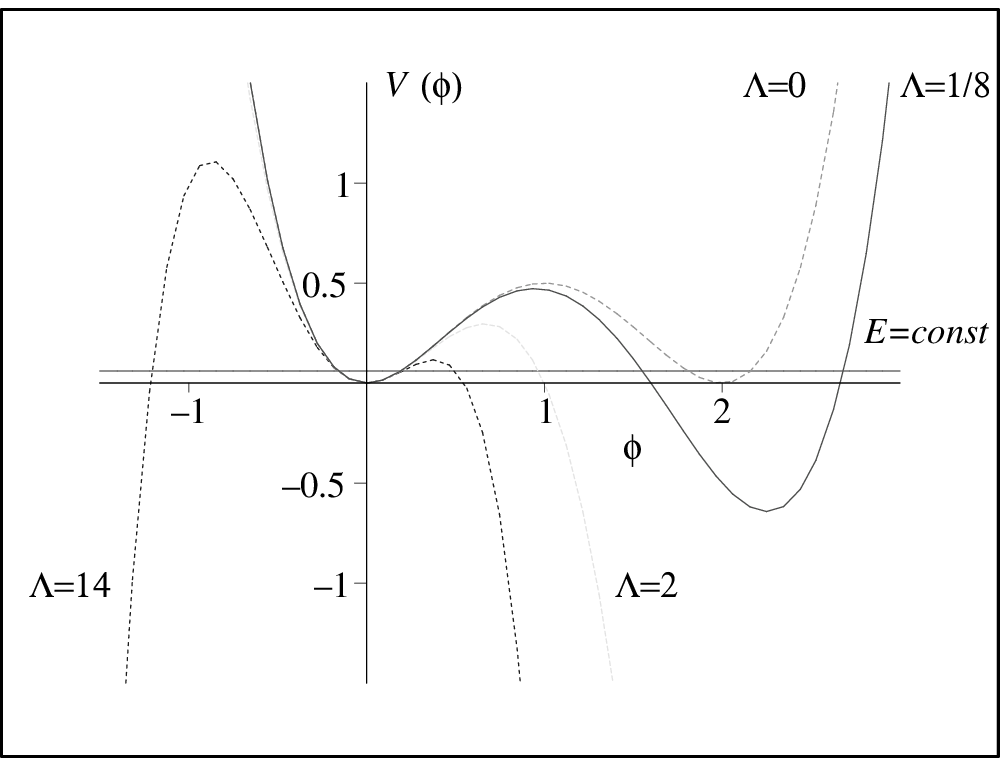}\\
  \vspace{3mm}
  \footnotesize {Fig. 1. Different shapes of the effective potential \\
  for different values of $\Lambda$.}
  \end{center}
The potential $V(\phi(\tau))$ is quartic except for $\Lambda=2$ in which
case it becomes cubic. For nonzero values of $\Lambda$ the potential is
asymmetric, indeed for $0< \Lambda <2$ there are two unequal  minima at
\begin{equation}
\phi_{1min}=0,\qquad \phi_{2min}=\frac{3+\sqrt{1+4\Lambda}}{2-\Lambda}>0
\end{equation}
and one  maximum at
\begin{equation}
\phi_{max}=\frac{3-\sqrt{1+4\Lambda}}{2-\Lambda}>0.
\end{equation}
 For $\Lambda>2$ the shape of the potential is changed
  and there are  two maxima at
\begin{equation}
\phi_{lmax}= \frac{-3-\sqrt{1+4\Lambda}}{\Lambda-2}<0,\;
\phi_{smax}=\frac{-3+\sqrt{1+4\Lambda}}{\Lambda-2}>0,
\end{equation}
with the small  and large barriers $V_{smax}$ and $V_{lmax}$ and one
minimum at
\begin{equation}
\phi_{min}=0.
\end{equation}
In the case of $\Lambda=2$  the potential is cubic  with the minimum at
$\phi_c=0$ and maximum at   $\phi_c=2/3$ . The case of $\Lambda=0$ is just
the gauge theory with local $SU(2)$-symmetry and the  effective potential
is symmetric in this case with two minima at $\phi=0,2$  and
 one maximum at $\phi=1$. All these cases are shown in Fig. 1.

The points
 of intersection of the line $E=const$ with the potential determine the
 turning points of the periodic motion.

\section {Quantum-classical transitions.}

\subsection {The case $0<\Lambda<2$}

The equation of motion for nonzero constant of integration $E$ reads:
\begin{equation}
\frac{1}{2}\left(\frac{\partial \phi}{\partial
\tau}\right)^2=V(\phi(\tau))-E.
\end{equation}
The solution, which satisfies the periodicity condition $\phi(\tau
+P(E))=\phi(\tau)$ is found to be:
\begin{equation}
\phi(\tau)=\frac{c(b-d)-d(b-c){\rm sn}^{2}(B(\Lambda,E)\tau + {\rm
K}(k),k)}{(b-d)-(b-c){\rm sn}^{2}(B(\Lambda,E)\tau +{\rm K}(k),k)},
\end{equation}
in which
\begin{equation}
B(\Lambda,E)=\sqrt{\frac{(2-\Lambda)}{8}}\sqrt{(b-d)(a-c)},
\end{equation}
and $k$ is the modulus of the Jacobian elliptic function
\begin{equation}
k=\sqrt{\frac{(b-c)(a-d)}{(a-c)(b-d)}}.
\end{equation}
The function ${\rm K}(k)$ is the complete elliptic integral of the first
kind. The quantities $a, \; b, \; c,\; d$ with  $a>b>c>d$ are turning
points of the motion (see Appendix). The solution corresponds to the
periodic motion from the point $c$ via the maximum of the potential to $b$
and back. The period of motion is
\begin{equation}
P(E)=\frac{2{\rm K}(k)}{B(\Lambda,E)},
\end{equation}
so that
$$
\phi(-P(E)/2)=\phi(P(E)/2)=c \quad {\rm and} \quad \phi(0)=b.
$$
The period $P(E)$ is a monotonous function of the energy, as shown in Fig.
2. This indicates  that the quantum-classical phase transition is of
second order.
\begin{center}

  \includegraphics[width=8cm]{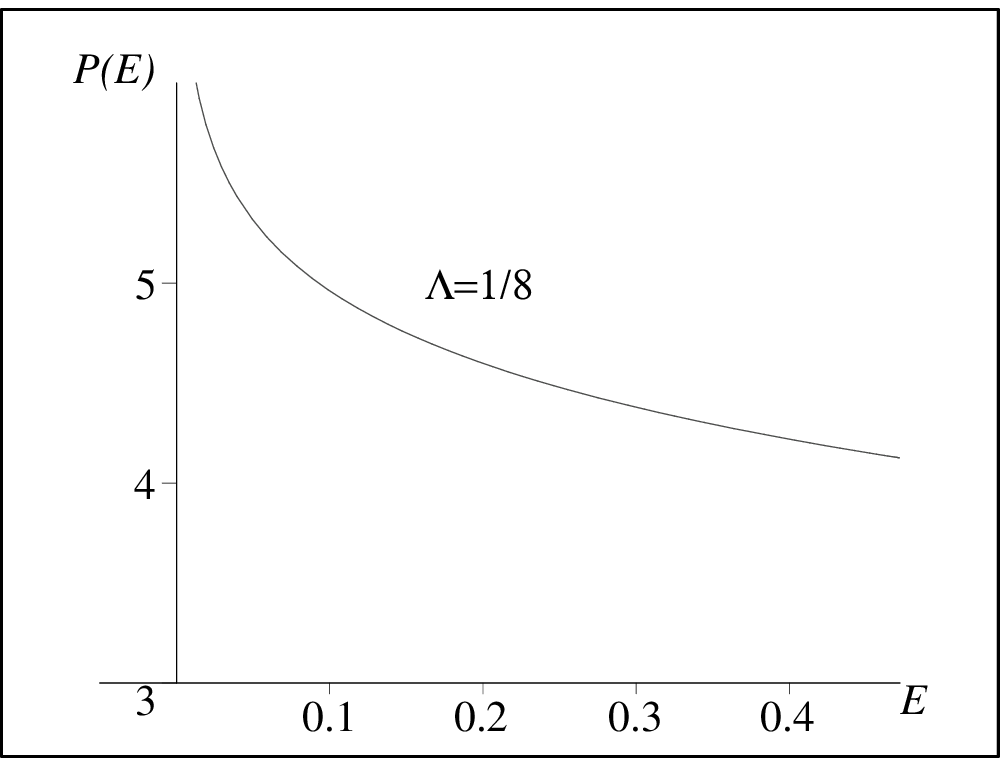}\\
  \vspace{3mm}
 \footnotesize {Fig. 2. Energy dependence of the period\\
  for $\Lambda=1/8$, which is typical for all $\Lambda$.}
\end{center}
Substituting the solution into the action and integrating over the period
gives the following expression:
\begin{eqnarray}
S(\Lambda,E)=\frac{6\pi^2}{g^2}\bigl\{E P(E)+D(\Lambda,E){\rm
K}(k)+G(\Lambda,E){\rm E}(k)+\nonumber \\
         H(\Lambda,E){\Pi}(\alpha^2,k)\biggr\},
\end{eqnarray}
with
$$
\alpha^2=\frac{b-c}{b-d}>0
$$
The  functions $D(E),\; G(E), \; H(E)$ are defined in  the Appendix. The
functions ${\rm K}(k),\; {\rm E}(k)$  and $\Pi(\alpha^2,k)$ are complete
elliptic integrals of the first, second and third kind respectively. In
semiclassical approximation  the solution describes the quantum tunneling
process. As the solution responsible for  the thermal activation we choose
the constant solution of the equation of  motion, namely
$\phi(\tau)=\phi_{max}$ which corresponds to the ``particle'' with the
maximal energy $E=V_{max}=V(\phi_{max})$ sitting at the top of the
potential barrier. The action of this configuration is
\begin{equation}
S_0(T)=\frac{6\pi^2}{g^2}\frac{V_{max}}{T},
\end{equation}
where $T$ is the  temperature. In Fig. 3 we display the
action-versus-temperature plot (with $P(E)=1/T(E)$ in (17)). One can see,
that the transition is of the second order with the transition temperature
$T_{ cr}=0.24$.

\begin{center}

  \includegraphics[width=8cm]{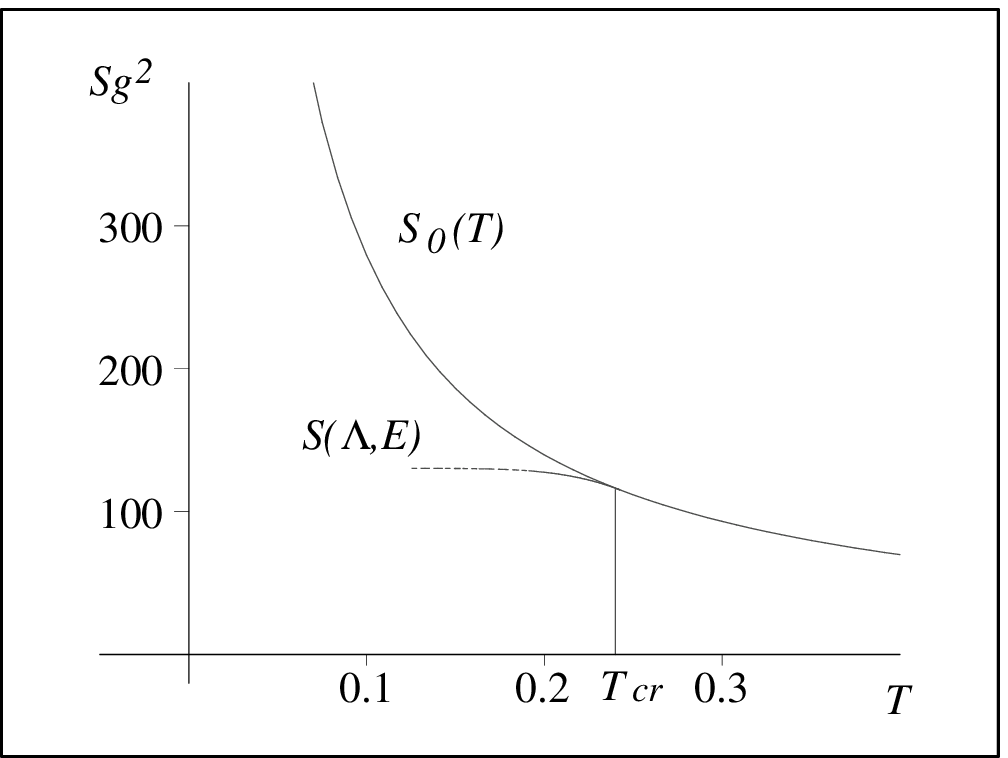}\\
\vspace{3mm}
 \footnotesize {Fig. 3.  The action-versus-temperature diagram:
the solid line \\represents the action
 of the the sphaleron
$\phi{(\tau)}=\phi_{max}$ \\ and the dashed line the action of the
periodic  instanton \\for $\Lambda=1/8$.}

\end{center}

The case of $\Lambda=0$ (theory with local symmetry) is included in our
formulae
  for $0<\Lambda<2$ by setting $\Lambda=0$. Nevertheless we give the exact expressions for the solution of the problem. The turning points
for the periodic motion with finite``energy'' as the solutions of the
equation
$$V(\phi)=\frac{1}{2}\phi^4-2\phi^3+\phi^2=E$$
 are:
$$
1-\sqrt{1+\sqrt{E}},\;1-\sqrt{1-\sqrt{E}},\;1+\sqrt{1-\sqrt{E}},\;1+\sqrt{1+\sqrt{E}}.
$$
The periodic field configuration, which describes the motion between the
points $1-\sqrt{1-\sqrt{E}}$   and $1-\sqrt{1+\sqrt{E}}$ is
\begin{equation}
\phi(\tau)=1+\sqrt{1-\sqrt{E}}{\rm sn}\sqrt{1+\sqrt{E}}(\tau+P_0(E)/2)
\end{equation}
and satisfies the condition $\phi(-P_0(E)/2)=\phi(P_0(E)/2)$, in which
$$P_0(E)=\frac{4}{\sqrt{1+\sqrt{E}}}{\rm K}(k_0)$$
is a period of motion with  modulus
$$k_0=\sqrt{\frac{1-\sqrt{E}}{1+\sqrt{E}}}.$$
The values of the actions corresponding to the periodic motion and the
static field configuration $\phi=1/2$ (at the top of the barrier)  are
then
$$
S_0(E)=\frac{6\pi^2}{g^2}\left\{\frac{E}{T(E)}+
\frac{4}{3}(1+\sqrt{E})\sqrt{1-\sqrt{E}}\left[(1+k_0^2){\rm
E}(k_0)-(1-k_0^2){\rm K}(k_0)\right]\right\},$$
$$S_{0st}=\frac{3\pi^2}{g^2T}.$$
with  temperature $T_0(E)=P^{-1}_0(E)$. The period is a
decreasing function of the energy and the phase transition is
of second order. Finally we  mention, that although we have
restricted ourselves to positive values of the parameter $\Lambda$,
some negative values may also be allowed, in particular one sees from
the expressions of the extrema of the potential $V(\phi(\tau))$ that
the maximum exists for $-1/4<\Lambda$. For the values of $\Lambda\leq-1/4$
the potential $V(\phi(\tau))$ does not have a maximum any more. \\

\subsection {The case $\Lambda>2$}

In the case of $\Lambda>2$ the ``particle'' sitting in the potential well
with minimal energy can move in the direction  of either a bigger  or
smaller barrier. The solutions of the field equations in both cases will
be presented in the form, which is different from that given by  (13). The
turning points $a_s>b_s>a_l>b_l$ are given by
$$
a_s=m_s+n_s,  \; b_s=m_s-n_s, \; a_l=m_l+n_l, \; b_l=m_l-n_l,\;
$$
where $=m_s,\; n_s, \; m_l, \; n_l$ are determined in the Appendix. We
consider first the motion to the small barrier. In this case the periodic
field configuration is:
\begin{equation}
\phi_s(\tau)=\frac{q_sb_s+p_sa_s-(q_sb_s-p_s a_s){\rm
cn}\bigl(\sqrt{\frac{(\Lambda-2)p_s
q_s}{2}}(\tau+P_s(E)/2),k_s\biggr)}{q_s+p_s-(q_s-p_s){\rm
cn}\bigl(\sqrt{\frac{(\Lambda-2)p_s q_s}{2}}(\tau+P_s(E)/2),k_s\biggr)},
\end{equation}
 with the conditions
\begin{eqnarray*}
\phi_s(-P_s(E)/2)&=&\phi_s(P_s(E)/2))=a,\\
\phi_s(-P_s(E)/4)&=&\phi_s(P_s(E)/4)=V_{\rm smax},  \\
 \phi_s(0)&=&b.  \\
\end{eqnarray*}
The real quantities $q_s,\;p_s$ and the modulus $k_s$ of the Jacobian
elliptic functions are defined as
\begin{eqnarray*}
q_s^2=(m_l-a_s)^2-n_l^2, \; p_s^2=(m_l-b_s)^2-n_l^2, \;
       k_s=\frac{1}{2}\sqrt{\frac{4n_s^2 -(p_s-q_s)^2}{p_sq_s}}. \\
\end{eqnarray*}
The period of the motion
\begin{equation}
P_s(E)= 4\sqrt{\frac{2}{(\Lambda-2)p_sq_s}}{\rm K}(k_s)
\end{equation}
is again a monotonically decreasing function of the energy $E$. The action
integrated out over the period  is a linear combination of complete
elliptic integrals:
\begin{eqnarray}
S_s(\Lambda,E)=\frac{6\pi^2}{g^2}\bigl\{E P_s(E)+D_s(\Lambda,E){\rm K}(k_s)+G_s(\Lambda,E){\rm E}(k_s)+\biggr.\nonumber \\
\biggl.H_s(\Lambda,E){\Pi}(\frac{a_s^2}{\alpha_s^2-1},k_s)\biggr\},
\end{eqnarray}
with $$\alpha_s^2=\frac{p_s-q_s}{p_s+q_s}.$$ The coefficients $D_s(E),\;
G_s(E), \; H_s(E)$) are given in the Appendix. Fig. 4 a) shows  the
action-versus-temperature plot, in which $$S_{s0}(T)=
\frac{6\pi^2}{g^2}\frac{V_{smax}}{T}$$ is an action corresponding to the
top of the small barrier.
\begin{center}

\includegraphics[width=6cm]{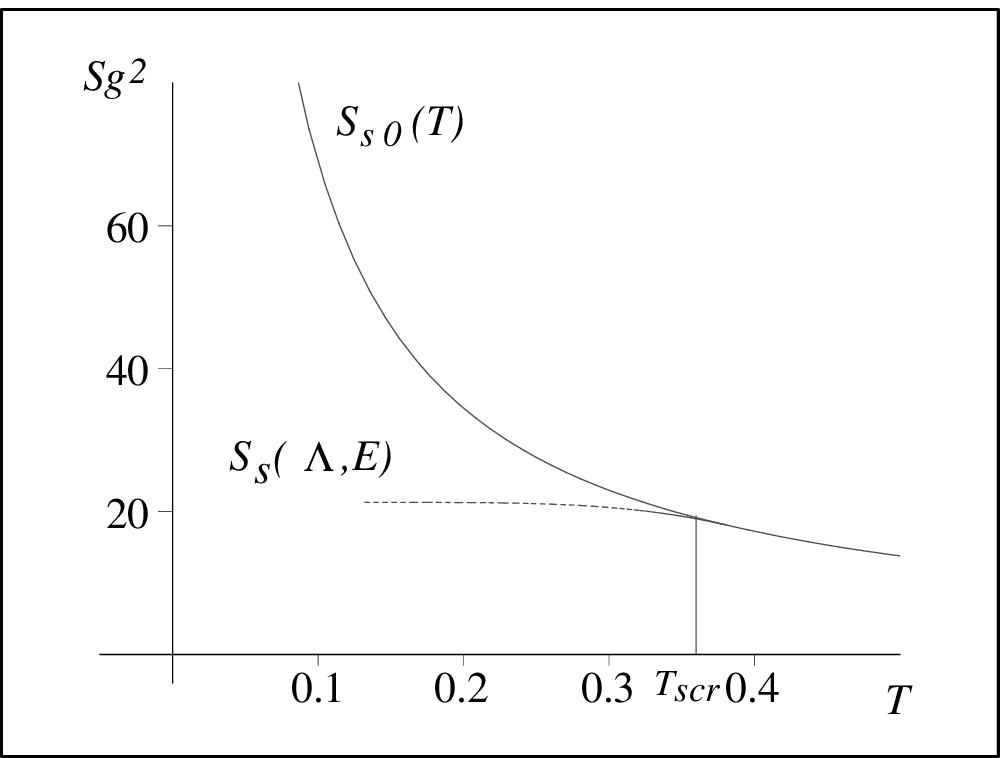}  \includegraphics[width=6cm]{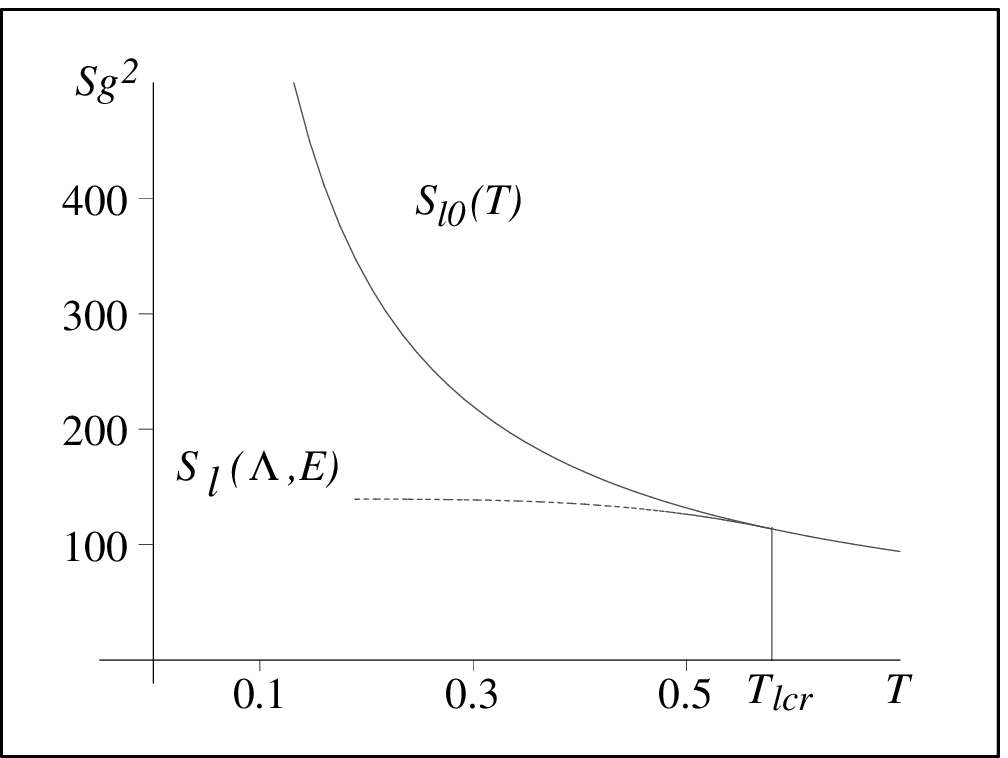}
\begin{flushleft}
a)\hspace{60mm}b)
\end{flushleft}
 \vspace{2mm}

 \footnotesize {Fig. 4. The action-versus-temperature diagram: a)For the small  barrier
the solid line represents the action $S_{s0}$ of the the sphaleron
$\phi{(\tau)}=\phi_{smax}$  and the dashed line the action
$S_s(\Lambda,E)$ of the periodic  instanton at $\Lambda=14$. b)For the
large barrier the solid line represents the action $S_{l0}$ of the the
sphaleron $\phi{(\tau)}=\phi_{lmax}$  and the dashed line the action
$S_l(\Lambda,E)$ of the periodic  instanton at $\Lambda=14$.}

\end{center}

In the case of motion to the large barrier the solution is defined by
formula (20) with appropriately  replaced coefficients, in particular
\begin{equation}
\phi_l(\tau)=\frac{q_lb_l+p_la_l-(q_lb_l-p_la_l){\rm
cn}\bigl(\sqrt{\frac{(\Lambda-2)p_lq_l}{2}}(\tau+P_l(E)/2),k_l\biggr)}{q_l+p_l-(q_l-p_l){\rm
cn}\bigl(\sqrt{\frac{(\Lambda-2)p_l q_l}{2}}(\tau+P_l(E)/2),k_l\biggr)},
\end{equation}
This solution  corresponds to  periodic motion starting at the point
$a_l$, going to $b_l$ and back to $a_l$. The quantities $p_l, \,q_l$ are
given by
$$q_l^2=(m_s-a_l)^2-n_s^2, \qquad   p_l^2=(m_s-b_l)^2-n_s^2.$$
The period $P_l(E)$ is a monotonically  decreasing function of $E$ and is
given by (21) with $k_s$  replaced as
$$k_s\rightarrow k_l=\frac{1}{2}\sqrt{\frac{4n_l^2-(p_l-q_l)^2}{p_lq_l}}.$$
The action $S_l(\Lambda,E)$ is also defined by (22) with different
coefficients $D_l(E), G_l(E),\\
 H_l(E)$ (see Appendix) and the parameter
$$\alpha_l^2=\frac{p_l-q_l}{p_l+q_l}$$ instead of $\alpha_s^2$. The action of the
sphaleron in this case is expressed through the value $V_{lmax}$:
$$S_{l0}=\frac{6\pi^2}{g^2}\frac{V_{lmax}}{T}.$$  The corresponding diagrams, again
 showing the existence of  the smooth second order phase transition,
are shown in Fig. 4 b). The critical temperatures are $T_{ lcr}=0.58$ and $T_{ scr}=0.38$.
 One can see, that tunneling through the large potential barrier dominates that of the
 small
 barrier ($S_{l}(\Lambda,E)>S_{s}(\Lambda,E)$ for a given ``energy'' $E$). With increase
  of the parameter $\Lambda$ this difference is diminished.

\subsection {The case $\Lambda=2$.}

In the case of  $\Lambda=2$ the finite ``energy'' periodic solution of the
equation of  motion is
\begin{equation}
\phi_c(\tau)=\frac{b_c(a_c-d_c)-d_c(a_c-b_c){\rm
sn}^2(\sqrt{\frac{a_c-d_c}{2}}(\tau+
\frac{P(E)}{2}))}{(a_c-d_c)-(a_c-b_c){\rm
sn}^2(\sqrt{\frac{a_c-d_c}{2}}(\tau+\frac{P(E)}{2}))},
\end{equation}
in which  $$k_c=\sqrt{\frac{a_c-b_c}{a_c-d_c}}$$ is the modulus of the
elliptic functions and the quantities $a_c>b_c>d_c$ are turning points
given by  the solutions of the cubic equation
\begin{equation}
-2\phi^3+2\phi^2=E.
\end{equation}
 The solution describes a motion from the point $b_c$ to $a_c$ and back.
 The period of motion $$P(E)=\frac{2\sqrt{2}{\rm K}(k_c)}{\sqrt{a_c-d_c}}$$
  is a monotonic function of $E$.
The  action integrated out over the period  reads:
\begin{eqnarray}
S_c(\Lambda,E)=\frac{6\pi^2}{g^2}\left\{\frac{E}{T(E)}\right.+
\frac{8(b_c-d_c)^2}{5}\sqrt{\frac{a_c-d_c}{2}}
\left[\left(\frac{2(2-k_c^2)}{3(1-k_c^2)^2}  \right)\right.  -\nonumber \\
\frac{2}{1-k_c^2}{\rm E}(k_c)
 \left.\left.-\frac{(2-k_c^2)}{3(1-k_c^2)}{\rm K}(k_c)\right]\right\}.
\end{eqnarray}
The maximum of the potential is $8/27$ and thus the action of the  static
field $\phi_c$ is $S_{c0}=16\pi^2/9T$.  The``energy'' dependence of the
period and the action-versus-temperature diagrams confirm the existence of
the phase transitions of the second order and  we do not   reproduce the
corresponding diagrams.

\section {Conclusions}

Above we demonstrated that in the  model considered the smooth phase
transitions of the second order take place. This is not surprising, since
it is known from general arguments  that in theories with cubic and
quartic terms the phase transitions of the second order occur. Although
our considerations are  explicit they are  based on the Ansatz (5), which
singles out a class of periodic field configurations  (which are
nonselfdual) reducing the problem to the one-dimensional one. In order to
check the criteria for  occurrence of a transition of the first
order\cite{Mue} one has to investigate fluctuations around a  static (in
our model  constant)  configuration. The nonselfdual character of the
solutions complicates the  second order fluctuation differential equations
for derivation of analytical solutions considerably, so that this is not
attempted here.

{\bf Acknowledgment.} J.-Q.L.and D. K. Park are indebted to The Deutsche
Forschungsgemeinschaft (Germany) for financial support of visits to
Kaiserslautern.

\begin{center}
{\large \bf Appendix }
\end{center}
We here  give some explicit formulae referred to above for the  $3$
possible domains of the parameter $\Lambda$.

a) $0<\Lambda<2.$

The turning points of the periodic motion as  solutions of the equation
 $$\frac{(2-\Lambda)}{4}\phi^4-2\phi^3+\phi^2=E$$ are:
\begin{eqnarray*}
a&=&\frac{1}{2}(r+\sqrt{r^2-2r+2y})+\frac{1}{2}\sqrt{2(r^2-r-y)+2r\frac{r^2-2r}{\sqrt{r^2-2r+2y}}},  \\
b&=&\frac{1}{2}(r+\sqrt{r^2-2r+2y})-\frac{1}{2}\sqrt{2(r^2-r-y)+2r\frac{r^2-2r}{\sqrt{r^2-2r+2y}}},  \\
c&=&\frac{1}{2}(r-\sqrt{r^2-2r+2y})+\frac{1}{2}\sqrt{2(r^2-r-y)-2r\frac{r^2-2r}{\sqrt{r^2-2r+2y}}}, \\
d&=&\frac{1}{2}(r-\sqrt{r^2-2r+2y})-\frac{1}{2}\sqrt{2(r^2-r-y)-2r\frac{r^2-2r}{\sqrt{r^2-2r+2y}}},
\
\end{eqnarray*}
in which
$$r=\frac{4}{2-\Lambda},$$
$$y=\frac{1}{3}r+2\sqrt{\frac{1}{3}\left(\frac{r^2}{3}-rE\right)}\cos\left(\frac{1}{3}
\arctan\frac{2\sqrt{-\frac{r^3}{27}E(E-V_{max})(E-V_{2min})}}{-\frac{2}{27}r^3+\frac{1}{2}r^2(r-\frac{4}{3})E}\right).$$
One  checks, that the quantities $a,\;b,\;c,\;d$ are real for ``energies''
$E$ in  $0<E<V_{max}$. The coefficients in the expression for the action
$S(\Lambda,E)$ are defined as
$$D(\Lambda,E)=\frac{4B(\Lambda,E)}{\alpha^2}(c-d)^2\left(\frac{\alpha^2k^2+\alpha^2-3k^2}{3(1-\alpha^2)}+
\frac{(2\alpha^2k^2+2\alpha^2-\alpha^4-3k^2)^2}{8(1-\alpha^2)^2(k^2-\alpha^2)}\right.$$
$$+\left.\frac{k^2(2\alpha^2k^2+2\alpha^2-\alpha^4-3k^2)}{12(1-\alpha^2)(k^2-\alpha^2)}\right),$$

$$G(\Lambda,E)=\frac{4B(\Lambda,E)(c-d)^2}{k^2-\alpha^2}\left(\frac{\alpha^2k^2+\alpha^2-3k^2}{3(1-\alpha^2)}+
\frac{(2\alpha^2k^2+2\alpha^2-\alpha^4-3k^2)^2}{8(1-\alpha^2)^2(k^2-\alpha^2)}\right),$$
$$H(\Lambda,E)=\frac{4B(\Lambda,E)(c-d)^2}{\alpha^2}\left(-k^2+\right.$$
$$\left.\frac{(\alpha^2k^2+\alpha^2-3k^2)(2\alpha^2k^2+2\alpha^2-\alpha^4-3k^2)}{2(1-\alpha^2)(k^2-\alpha^2)}+
\frac{(2\alpha^2k^2+2\alpha^2-\alpha^4-3k^2)^3}{8(1-\alpha^2)^2(k^2-\alpha^2)^2}\right).$$

b) $\Lambda>2.$

The quantities $m_s,\; n_s,\; m_l, \; n_l$ which determine the turning
points in this case, are given by
\begin{eqnarray*}
m_s&=&\frac{1}{2}(-r+\sqrt{r^2+2r+2y}), \nonumber \\
n_s&=&\frac{1}{2}\sqrt{4r(r+1)-(r-\sqrt{r^2+2r+2y})^2+4\sqrt{y^2-rE}}, \nonumber \\
m_l&=&\frac{1}{2}(-r-\sqrt{r^2+2r+2y}), \nonumber \\
n_l&=&\frac{1}{2}\sqrt{4r(r+1)-(r+\sqrt{r^2+2r+2y})^2-4\sqrt{y^2-rE}},
\nonumber
\end{eqnarray*}
with $$r=\frac{4}{\Lambda-2}$$ and
$$y=-\frac{1}{3}r+2\sqrt{\frac{1}{3}\left(\frac{r^2}{3}+rE\right)}\cos\left(\frac{1}{3}
\arctan\frac{2\sqrt{\frac{r^3}{27}E(E-V_{smax})(E-V_{\rm
lmax})}}{\frac{2}{27}r^3- \frac{1}{2} r^2(r+\frac{4}{3})E}\right).$$ The
coefficients which appear in the action $S_s$  read:
$$D_s(\Lambda,E)=\frac{8p_s^2q_s^2(a_s-b_s)^2}{3\alpha^4_s(p_s+q_s)^4}\sqrt{\frac{(\Lambda-2)p_sq_s}{2}
}\left[-4k_s^2-\frac{2(6k_s^2+\alpha_s^2-2\alpha_s^2k_s^2)}{(\alpha_s^2-1)(\alpha_s^2+
k_s^2-\alpha_s^2k_s^2)}\right.$$
$$-\left.\frac{3(2\alpha_s^2k_s^2-\alpha_s^2-2k_s^2)^2}{(\alpha_s^2-1)^2(\alpha_s^2+
k_s^2-\alpha_s^2k_s^2)^2}+\frac{2k_s^2(2\alpha_s^2k_s^2-\alpha_s^2-2k_s^2)}{(\alpha_s^2-1)(\alpha_s^2+
k_s^2-\alpha_s^2k_s^2)}
\right],$$
$$G_s(\Lambda,E)=\frac{8p_s^2q_s^2(a_s-b_s)^2}{3\alpha^4_s(p_s+q_s)^4}\sqrt{\frac{(\Lambda-2)p_sq_s}{2}
}\left[\frac{2\alpha_s^2(6k_s^2+\alpha_s^2-2\alpha_s^2k_s^2)}{(\alpha_s^2-1)(\alpha_s^2+k_s^2-\alpha_s^2k_s^2)}+
\right.$$
$$\left.\frac{3\alpha_s^2(2\alpha_s^2k_s^2-\alpha_s^2-2k_s^2)}{(\alpha_s^2-1)^2(\alpha_s^2+k_s^2-\alpha_s^2k_s^2)^2}\right],$$
$$H_s(\Lambda,E)=\frac{8p_s^2q_s^2(a_s-b_s)^2}{3\alpha^4_s(p_s+q_s)^4}\sqrt{\frac{(\Lambda-2)p_sq_s}{2}
}\left(\frac{12k_s^2}{1-\alpha_s^2}\right. +$$
$$\left.\frac{3(2\alpha_s^2k_s^2-\alpha_s^2-2k_s^2)^3}{(\alpha_s^2-1)^3(\alpha_s^2+k_s^2-\alpha_s^2k_s^2)^2}+
\frac{3(2\alpha_s^2k_s^2-\alpha_s^2-2k_s^2)(6k_s^2+\alpha_s^2-2\alpha_s^2k_s^2)}{(\alpha_s^2-1)^2(\alpha_s^2+k_s^2-\alpha_s^2k_s^2)}\right).$$
The corresponding quantities in $S_l(\Lambda,E)$ are  obtained by the following replacements:
$$a_s,\;b_s\rightarrow a_l,\;b_l,\; \alpha_s \rightarrow \alpha_l\,\; k_s\rightarrow k_l,\; p_s,\;q_s\rightarrow p_l,\;q_l.$$

c) $\Lambda=2.$

The turning points in this case are the solutions of the cubic equation
(25), in particular:
$$
a_c=\frac{1}{3}+\cos\frac{\phi}{3},\;
b_c=\frac{1}{3}+\cos\frac{\phi+2\pi}{3},\;
d_c=\frac{1}{3}+\cos\frac{\phi+4\pi}{3},$$ in which the angle ${\phi}$ is
defined by
$$\tan\frac{\phi}{3}=-\frac{2\sqrt{-\frac{1}{4}(-\frac{2}{27}+
\frac{E}{2})^2+\frac{1}{27}}}{-\frac{2}{27}+\frac{E}{2}}.$$

\end{document}